\documentclass[a4paper,twoside]{article}

\usepackage{epsfig}
\usepackage{subcaption}
\usepackage{calc}
\usepackage{amssymb}
\usepackage{amstext}
\usepackage{amsmath}
\usepackage{amsthm}
\usepackage{multicol}
\usepackage{pslatex}
\usepackage{apalike}
\usepackage{algorithm2e}
\usepackage[bottom]{footmisc}
\usepackage{blindtext}
\graphicspath{{gfx/}}
\usepackage{hyperref}
\usepackage{tikz}
\tikzset{slice/.append style={line width=1.5pt}}
\def\biglayersep{1.6cm}
\def\layersep{1.5cm}
\usepackage{subcaption}
\usepackage{orcidlink}

\usepackage{SCITEPRESS}     

\begin{document}

\title{Exploring Unsupervised Anomaly Detection with Quantum Boltzmann Machines in Fraud Detection}

\author{\authorname{Jonas Stein\sup{1}\orcidlink{0000-0001-5727-9151}, Daniëlle Schumann\sup{1}\orcidlink{0009-0000-0069-5517}, Magdalena Benkard\sup{1}\orcidlink{0009-0009-5461-815X}, Thomas Holger\sup{1}, Wanja Sajko\sup{1}, Michael Kölle\sup{1}\orcidlink{0000-0002-8472-9944}, Jonas Nüßlein\sup{1}\orcidlink{0000-0001-7129-1237}, Leo Sünkel\sup{1}, Olivier Salomon\sup{2} and Claudia Linnhoff-Popien\sup{1}\orcidlink{0000-0001-6284-9286}} 
\affiliation{\sup{1}LMU Munich, Germany}
\affiliation{\sup{2}Allianz France}
\email{jonas.stein@ifi.lmu.de}
}


\abstract{Anomaly detection in Endpoint Detection and Response (EDR) is a critical task in cybersecurity programs of large companies. With rapidly growing amounts of data and the omnipresence of zero-day attacks, manual and rule-based detection techniques are no longer eligible in practice. While classical machine learning approaches to this problem exist, they frequently show unsatisfactory performance in differentiating malicious from benign anomalies. A promising approach to attain superior generalization than currently employed machine learning techniques are quantum generative models. Allowing for the largest representation of data on available quantum hardware, we investigate Quantum Annealing based Quantum Boltzmann Machines (QBMs) for the given problem. We contribute the first fully unsupervised approach for the problem of anomaly detection using QBMs and evaluate its performance on an EDR inspired synthetic dataset. Our results indicate that QBMs can outperform their classical analog (i.e., Restricted Boltzmann Machines) in terms of result quality and training steps in special cases. When employing Quantum Annealers from D-Wave Systems, we conclude that either more accurate classical simulators or substantially more QPU time is needed to conduct the necessary hyperparameter optimization allowing to replicate our simulation results on quantum hardware.
} 

\onecolumn \maketitle \normalsize \setcounter{footnote}{0} \vfill

\section{\uppercase{Introduction}}
\label{sec:introduction}
Anomaly detection is the identification of abnormal behavior in data, which manifests in individual data points that differ significantly from the majority of the data~\cite{10.1145/1541880.1541882}. This task frequently appears in many domains including finance, medicine and telecommunications~\cite{618940,991693,1541896}. A particularly challenging application of Anomaly Detection can be found in Endpoint Detection and Response (EDR), which aims at detecting and investigating suspicious activities on endpoints such as mobile phones or workstations in cybersecurity~\cite{Chuvakin13}. In practice, the respective networks can be comprised of billions of nodes, generating an immense amount of data, in which the search for extremely scarce, malicious anomalies can be very tedious.

This vast number of typically high-dimensional data points and additional impediments such as zero-day attacks raise a demand for suitable anomaly detection techniques deviating from the still widely-used manual and rule-based approaches. While many classical machine learning approaches to this mostly unsupervised learning problem exist, (e.g., clustering~\cite{MUNIYANDI2012174}, autoencoders~\cite{Finke2021} or Bayesian networks~\cite{MASCARO201484}), the distinction between malicious and benign anomalies frequently remains intractable due to insufficient generalization~\cite{KARAMI201836}. This problem substantiates in a trade-off between detecting an unacceptably high number of false positives (i.e., benign data) and failing to reliably detect all true positives (i.e., the malicious anomalies).

In search for alternative approaches that can cope with the encountered real world data better, we investigate the application of Quantum Computing (QC) to this problem, as QC has shown promising performance in generative data modelling, which is a popular technique used in unsupervised anomaly detection~\cite{PhysRevResearch.4.043092}. The general motivation behind using generative models for unsupervised anomaly detection is that their core functionality of replicating a given data distribution yields a data model that closely represents the input dataset with the exclusion of all anomalies, as they are too scarce to be learned reliably. Therefore, anomalies can be detected by verifying their absence in the generated data distribution. Two possible advantages in using quantum generative models over their classical analogs are likely that they need less data points in training and that they can efficiently sample from specific classically intractable data distributions~\cite{Caro2022,hangleiter2023computational}. 

A particularly powerful type of quantum generative models are Quantum Boltzmann Machines (QBMs)~\cite{Amin18}, which have been shown to be universal approximators for probability distributions~\cite{194417,YOUNES1996109,Amin18,wiebe2019hidden}. QBMs can be understood as a quantum analog to classical Boltzmann Machines. The core difference is displayed in the sampling procedure, in which the QBM represents an approximate Boltzmann distribution in a quantum state, which generally allows for efficient sampling via measurements. Curious to investigate the scaling performance of such quantum approaches in terms of, i.a., the dimensionality of the dataset, we select a Quantum Annealing based implementation of this sampling step, as its physical implementations have a key advantage over the usage of gate based quantum computers: They currently offer the highest number of qubits, which is the central factor for the representable dimensionality of the dataset.

Framed by related concepts in literature, we contribute the first fully unsupervised anomaly detection based on QBMs and evaluate its performance on suitably generated synthetic datasets. Our core contributions to the investigation of the applicability of Quantum Boltzmann Machines for unsupervised anomaly detection amount to the following:
\begin{itemize}
    \item We introduce the first fully unsupervised anomaly detection approach using QBMs.
    \item We conduct a case study evaluating the presented approach on two D-Wave Quantum Annealers while using a similarly sized classical Restriced Boltzmann Machine (RBM) as baseline.
\end{itemize}

The subsequent contents of this paper are structured as follows. In section~\ref{sec:background}, we explain the functionality of Boltzmann Machines and their variants. In section~\ref{sec:relatedwork}, we give an overview of related work. In section~\ref{sec:methodology}, we present our methodology on how Quantum Boltzmann Machines can be used to detect anomalies in a fully unsupervised manner. In section~\ref{sec:experiments}, we show how suitable hyperparameters can be chosen and evaluate the results achieved, comparing classical and quantum hardware. Finally, section~\ref{sec:conclusion} concludes our findings and demonstrates possible future work.

\section{\uppercase{Background}}
\label{sec:background}
\subsection{Boltzmann Machines} \label{sec:BM}
A classical Boltzmann Machine (BM) is an undirected, stochastic neuronal network, which typically consists of two or more separate layers and is not restricted in its nodes' connections. For the sake of simplicity, we assume a very basic QBM consisting of two layers in the following explanation. BMs contain exactly one input layer which simultaneously acts as the output layer and is also referred to as \emph{visible} layer $v = (v_1, ... , v_N) \in \{0,1\}^N$, consisting of \emph{visible} units $v_i$. The other layers are referred to as \emph{hidden} layers $h = (h_1, ... , h_M) \in \{0,1\}^M$ and likewise consist of \emph{hidden} units $h_i$. Both hidden and visible units are restricted to assume the values zero or one. The probability of a specific configuration of $(v,h)$ occurring in practice is determined by a stochastic distribution -- the Boltzmann distribution $p$ \cite{fischer2012introduction} in which $T>0$ denotes a specific parameter called temperature, which we can assume as a given constant for now:
\begin{align*}
    p(v,h,\theta) &= \dfrac{\exp\left(-\dfrac{E(v,h,\theta)}{T}\right)}{Z} \\ \text{with } Z &= \sum_{\left(v,h\right)\in\{0,1\}^{N\times M}} \exp\left(-\dfrac{E(v,h,\theta)}{T}\right)
\end{align*}
\noindent where $\theta \equiv \{ W_{ij}, b^{(0)}_{i}, b^{(1)}_{j} \}$ denotes the weights $W_{ij}$ between units as well as the biases and acting on all visible and hidden units $b^{(0)}_{i}$ and $ b^{(1)}_{j}$. The Boltzmann distribution is determined by the energy function $E$, which is generally expressed as an Ising Hamiltonian:
\begin{align} \label{eq:bm-energiefunktion}
    E(v,h,\theta) =& - \sum^N_{i=1}\sum^M_{j=1} W_{ij} v_i h_j - \sum^N_{i=1}\sum_{k<i} W_{ik} v_i v_k \nonumber\\
    &- \sum^M_{j=1}\sum_{l<j} W_{jl} h_j h_l - \sum^N_{i=1} b^{(0)}_{i} v_i - \sum^M_{j=1}  b^{(1)}_{j} h_j
\end{align}
Choosing the \emph{Kullback-Leibler divergence} (KL divergence) as the corresponding loss function and combining it with training methods such as stochastic gradient descent, BMs can be trained so that their Boltzmann distributions match the distribution of a given dataset in a straightforward manner. The KL divergence is a measure of similarity between two probability distributions, which becomes zero when the two distributions are identical and yields higher values the more dissimilar the distributions are. Its gradient, which is calculated repeatedly throughout the BM's training process and conveniently takes the following simple form~\cite{ackley1985learning}:
\begin{align} \label{eq:kl-gradient}
    \frac{\partial D_{KL}\left(P_{\text{data}}\mid\mid P_{\text{model}}\right)}{\partial W_{ij}} =& \langle s_i s_j \rangle_{\text{data}} - \langle s_i s_j \rangle_{\text{model}} \\
    \frac{\partial D_{KL}\left(P_{\text{data}}\mid\mid P_{\text{model}}\right)}{\partial b^{(\cdot)}_{i}} =& \langle s_i  \rangle_{\text{data}} - \langle s_i \rangle_{\text{model}}
\end{align}
Here $W_{ij}$ is the weight associated with the connection between the neurons $s_i, s_j \in \{v_1,...,v_N, h_1,...,h_M\}$. $\langle s_i s_j \rangle$ denotes the expectation value of the product of the neuron values $ s_i$ and $ s_j $, i.e., the probability that both neurons assume the value one. Analogously, $\langle s_i \rangle$ denotes the probability of $s_i$ assuming the value one.

Computing the exact values for all partial derivatives generally demands calculating the energy of every possible combination of states $v\in \{0,1\}^N$, $h\in \{0,1\}^M$, making this approach intractable in practice as there are exponentially many in terms of the input domain $\{0,1\}^N$. Fortunately, sufficiently well approximating heuristic methods often exist that allow for a shorter runtime. The core idea for these approaches is repeated sampling from the respective distributions, represented by the BM, and averaging the results. Sampling is generally performed in two phases: First, in the \emph{clamped} phase, the values of the visible neurons are fixed to the bits of a randomly selected data point. In this phase, only the values $s_i$ of the hidden units, which now depend on the data point's value, are subsequently sampled from the BM's Boltzmann distribution. Together with the input data point, these values can be viewed as a sample from the data's probability distribution $P_{\text{data}}$ and used to calculate $\langle s_i \rangle_{\text{data}}$ and $\langle s_i s_j \rangle_{\text{data}}$. The expectation values $\langle s_i \rangle_{\text{model}}$ and $\langle s_i s_j \rangle_{\text{model}}$ for the model are then determined in the \emph{unclamped} phase, by sampling from the Boltzmann distribution associated with the BM using techniques like Markov chain Monte Carlo. These samples are subsequently used to calculate $\langle s_i \rangle_{\text{model}}$ and $\langle s_i s_j \rangle_{\text{model}}$. In practice, sampling from BMs is hence typically performed by iteratively computing the values of each neuron, depending on the values of its neighboring neurons, until an equilibrium is reached. As this has to be done for each sample, while multiple samples have to be calculated for each data point in each training epoch, the training time quickly takes intractably long.~\cite{Amin18,ackley1985learning,Adachi15}

\subsection{Restricted Boltzmann Machines} \label{sec:RBM}
When using Restricted Boltzmann Machines (RBMs), considerably shorter training compared to the standard BMs can be achieved. RBMs restrict all possible neuron connections so that only interlayer weights can be non-zero, making it possible to sample from visible and hidden units separately. This allows for faster sampling, as hidden and visible units only depend on the neuron values in the opposing layer, which are known in the clamped phases.
For RBMs, the energy function is thus reduced to~\cite{fischer2012introduction}:
\begin{equation} \label{eq:rbm-energiefunktion}
    E(v,h,\theta) = \sum^N_{i=1}\sum^M_{j=1} W_{ij} v_i h_j - \sum^N_{i=1} v_i b^{(0)}_{i} - \sum^M_{j=1}h_j b^{(1)}_{j}
\end{equation}
As it remains intractable to calculate the gradient of the weights and biases directly, sampling from the given Boltzmann distribution is mandatory. However, even for the RBM, drawing independent samples from the model in order to approximate the gradient is computationally expensive. Even though approximate sampling techniques like Contrastive Divergence can often be used effectively, their trade-off in runtime against accuracy is frequently worse than state-of-the-art classical generative models besides BMs for big datasets.~\cite{gabrie2015training}


\subsection{Quantum Boltzmann Machines} \label{sec:QBM}
A very promising approach towards speeding up the time consuming sampling process in classical (restricted) BMs are Quantum Boltzmann Machines (QBMs), which use quantum algorithms to prepare a quantum state that resembles the desired probability distribution and allows for sampling from it via measurements. One such quantum algorithm is Quantum Annealing, which has been shown to be capable of approximating Boltzmann distributions when executed on analog quantum hardware natively implementing this algorithm, i.e., Quantum Annealers~\cite{Amin18}. Quantum Annealing conducts a time evolution starting in the ground state of a known Hamiltonian that gradually evolves into an Ising Hamiltonian $\hat{H}_P$ that is often used to model the cost landscape of an optimization problem. When this time evolution is done adiabatically (i.e., not too fast) and without any hardware errors, the final state is guaranteed to be the ground state of the Hamiltonian $\hat{H}_P$, i.e., the global minimum of the cost function~\cite{PhysRevE.58.5355}. When conducting this process on a physical Quantum Annealer however, the system naturally interacts with the environment, which interestingly perturbs the final state to approximate a classical Boltzmann distribution over the energy function described by the Hamiltonian $\hat{H}_P$. The temperature of the resulting approximate Boltzmann distribution is determined by device specific properties in correlation with the specific Ising Hamiltonian and can be tuned by rescaling the weights and biases by the inverse of the so-called effective temperature, which can be calculated efficiently as shown in~\cite{Benedetti16}.~\cite{Adachi15,heuristic,Benedetti16}

As Quantum Annealing drastically reduces the amount of steps needed to create a sample, a quantum advantage in the form of a speedup can be expected here. Experiments of \cite{Amin18} show that QBMs can achieve better KL divergences than the BMs employed in their study when given the same runtime capabilities, suggesting that QBMs can also outperform their classical analogs in terms of result quality. Another advantage of QBMs is that they do not require restricting the connectivity of the BMs architecture, allowing for more complex models with a higher number of connections~\cite{hinton2012practical,Adachi15}.

\section{\uppercase{Related Work}}
\label{sec:relatedwork}
Our contribution to existing literature is resembled by the first successful application of QBMs for fully unsupervised anomaly detection. This represents a novel use case for QBMs trained using unsupervised techniques extending the portfolio of known productive QBM applications like image reconstruction~\cite{Benedetti17} or image generation~\cite{sato2021assessment}.

In regards to supervised learning, QBMs have shown promising performance for anomaly detection in applications like the classification of cybersecurity data, for which \cite{dixit2021training} showed that their Quantum Annealing based RBM can provide similar results to comparable classical RBMs. \cite{mphasis} proposed a semi-supervised approach to anomaly detection for credit card transaction data that employs an ensemble of quantum-inspired RBMs. For sampling, they used a set of solvers from the “Azure Quantum-Inspired Optimization (QIO)” suite instead of real quantum algorithms, while excluding the anomalies from the training data. They subsequently calculate the energy values of all data points (i.e., including anomalies) for all RBMs of the ensemble analytically and then identify an energy threshold separating normal data from the anomalies.

Beyond QBMs, other quantum generative models have been shown to perform well on similar tasks: \cite{bermot2023quantum} haven shown the effective applicability of Quantum Generative Adversarial Networks for anomaly detection in high energy physics, \cite{PhysRevD.105.095004} used a Quantum Autoencoder for a very similar use case, and \cite{schuhmacher2023unravelling} applied a Quantum Support Vector Classifier to find beyond standard model physics in data recorded at the LHC.

\section{\uppercase{Methodology}}
\label{sec:methodology}


In line with known techniques to use generative models for anomaly detection, we utilize a QBM as a generative model and subsequently identify anomalies by their infrequence in the generated model~\cite{HOH2022629,Luo2022,10.1007/978-3-031-18576-2_8,Amin18}. More specifically, we classify a given data point as anomalous, if its energy value in the QBM exceeds a specific limit. To implement the presented approach, we now first specify the QBM model architecture and then choose an energy threshold separating normal from anomalous data.

\subsection{QBM model architecture}
\begin{figure}[t]
 \centering
  \begin{subfigure}{0.45\columnwidth}
  \scalebox{0.8}{
    \begin{tikzpicture}[shorten >=1pt,-,draw=black!50, node distance=\layersep]
    \tikzstyle{every pin edge}=[<-,shorten <=1pt]
    \tikzstyle{neuron}=[circle,fill=black!25,minimum size=17pt,inner sep=0pt]
    \tikzstyle{visible neuron}=[neuron, fill=blue!50];
    \tikzstyle{hidden neuron}=[neuron, fill=gray!50];
    \tikzstyle{annot} = [text width=4em, text centered]

    \foreach \name / \y in {1,...,4}
        \node[visible neuron] (V-\name) at (0,-\y) {};

    \foreach \name / \y in {1,...,7}
        \path[yshift=1.55cm]
        node[hidden neuron] (H-0-\name) at (\biglayersep,-\y cm) {};

    \foreach \source in {1,...,4}
        \foreach \dest in {1,...,7}
            \path (V-\source) edge (H-0-\dest);

    \node[annot,below of=H-0-7, node distance=1cm] (hl 0) {Hidden layer};
    \node[annot,left of=hl 0, node distance=2cm] (vli) {Visible layer};
    \end{tikzpicture}
}
  \caption{RBM}
  \label{fig:rbm}
  \end{subfigure}
  \hfill
  \begin{subfigure}{0.45\columnwidth}
  \scalebox{0.8}{
   \begin{tikzpicture}[shorten >=1pt,-,draw=black!50, node distance=\layersep]
    \tikzstyle{every pin edge}=[<-,shorten <=1pt]
    \tikzstyle{neuron}=[circle,fill=black!25,minimum size=17pt,inner sep=0pt]
    \tikzstyle{visible neuron}=[neuron, fill=blue!50];
    \tikzstyle{hidden neuron}=[neuron, fill=gray!50];
    \tikzstyle{annot} = [text width=4em, text centered]

    \foreach \name / \y in {1,...,4}
        \node[visible neuron] (V-\name) at (0,-\y) {};

    \path (V-1) edge (V-2);
    \path (V-1) edge [bend right = 30] (V-3);
    \path (V-1) edge [bend right = 50] (V-4);
    \path (V-2) edge (V-3);
    \path (V-2) edge [bend right = 30] (V-4);
    \path (V-3) edge (V-4);

    \foreach \name / \y in {1,...,7}
        \path[yshift=1.55cm]
        node[hidden neuron] (H-0-\name) at (\biglayersep,-\y cm) {};

    \foreach \source in {1,...,4}
        \foreach \dest in {1,...,7}
            \path (V-\source) edge (H-0-\dest);

    \node[annot,below of=H-0-7, node distance=1cm] (hl 0) {Hidden layer};
    \node[annot,left of=hl 0, node distance=2cm] (vli) {Visible layer};
    \end{tikzpicture}
}
  \caption{QBM} 
  \label{fig:qbm} 
  \end{subfigure} 
 \caption[Schematic architecture of the BMs used]{Schematic visualization of the architecture of the BMs used in this paper.}
 \label{fig:bms} 
\end{figure}
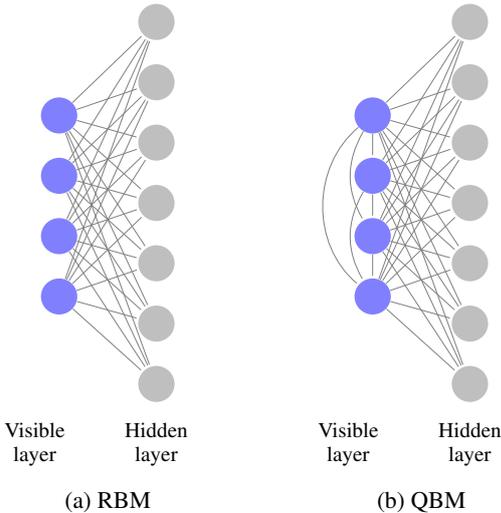

To perform anomaly detection in a fully unsupervised manner using a QBM, a suitable model architecture must be selected. Inspired from work of \cite{Amin18}, where a semi-restricted QBM (i.e., a QBM with one hidden layer, having lateral connection only between the visible neurons) was utilized for unsupervised learning, we too choose to also employ this model scheme. A general overview of this architecture can be found in figure~\ref{fig:qbm}. More specifically, we allow all neuron connections possible in this architecture and treat the number of hidden neurons as a hyperparameter, while the number of visible neurons is completely dependent on the dimensionality of the dataset.

\subsection{Choosing an Energy Threshold}
As stated previously, the identification of anomalies is done by verifying their absence (or at least significant infrequence) in the model distribution. For Energy Based Models (EBMs) like the QBM, the probability of a point in our model is fully dependent on its energy value by definition~\cite{928a56b7d6f1473e930f282a0c4b534e}. Thus, the classification of data points as anomalous or normal can be achieved by drawing an energy threshold between the two categories \cite{zhai2016deep,do2018energy}. This causes our model to classify all points with energy values higher than the threshold as anomalous and all points with lower energies as normal. While related work in the quantum domain specified this threshold for anomaly detection using (semi-)supervised learning~\cite{mphasis}, we draw the threshold in an unsupervised manner, by having the Quantum Annealer return the energy values of all data points in the training dataset 
and subsequently calculating the threshold as the $p$-th percentile, analog to in~\cite{zhai2016deep} and~\cite{do2018energy}. For our evaluation in section~\ref{sec:experiments}, we chose $p = 95$, analog to~\cite{do2018energy}, assuming that no more than $5\%$ of the training data is anomalous. 
While this amount is unrealistically high for the data employed in our use case, we choose it to increase the likelihood of finding all anomalies in the test dataset. 
If any kind of ad-hoc testing for the validity of an anomaly is available, e.g., through human inspection as is typically the case for EDR, this percentile can be tuned accordingly (see, e.g., \cite{do2018energy,zhai2016deep}).

\section{\uppercase{Evaluation}}
\label{sec:experiments}

To give an indication whether our proposed method might provide a quantum advantage over purely classical EBM-based approaches, we compare our approach to a RBM with a similarly simple architecture to our semi-restricted QBM: The chosen RBM is described in section~\ref{sec:RBM} and displayed in figure~\ref{fig:rbm}. As we treat the size of the hidden layer as a hyperparameter and as one hidden layer already suffices for the RBM to be an arbitrary distribution approximator, this choice yields a potent, practical, classical baseline. Preparing for our experiments, we now select a suitable dataset and conduct an in-depth hyperparameter optimization. Note that we use a $50/50$ train/test split in this evaluation and only show the results from test data. Furthermore, we averaged over three seeds in the hyperparameter optimization and over five in the final evaluation, due to severe runtime demands.

\subsection{Dataset}
\begin{figure*}[t]
    \centering
    \includegraphics[width=\textwidth]{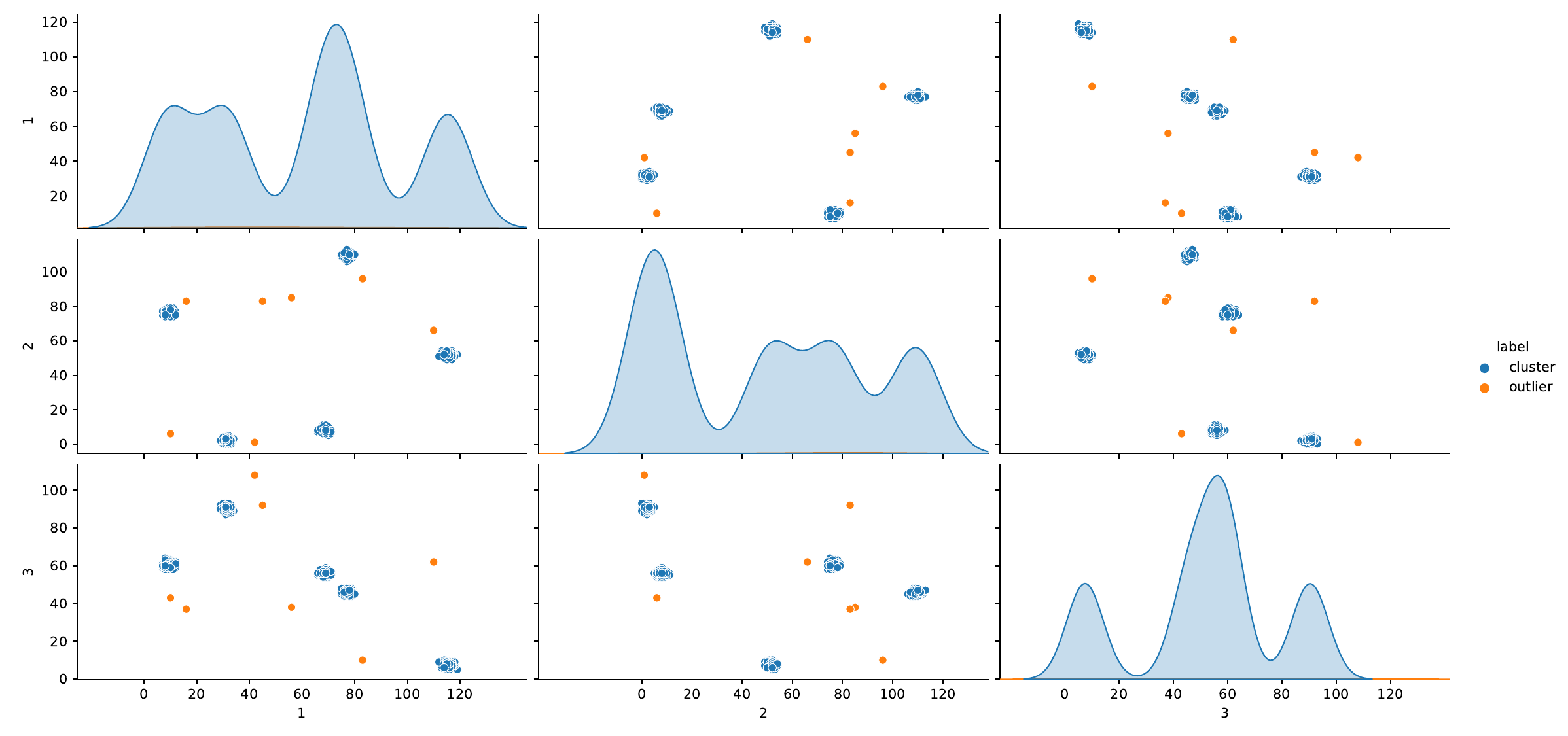}
    \caption{Visualization of the constructed 3D dataset comprised of 5 clusters and 7 anomalies. The diagonal shows the distribution of the data in each dimension. The off-diagonal plots, show a 2D flattened version of the dataset for all dimension-combinations.}
    \label{fig:dataset}
\end{figure*}

In lack of suitably small real world EDR datasets that fit on current quantum hardware, we generate a synthetic data set, aimed at matching the following properties found in real world data: (1) a high dimensionality to assess scaling performance, (2) scarce anomalies and (3) a substantial number of data points. While satisfying (2) and (3) is straightforward, (1) is direct proportional to the number of visible units and thus the space complexity compromising on the dimensionality. Compromising (1) to facilitate a visual evaluation and retain enough space for exploring a large space of hidden units for this first case study on fully unsupervised anomaly detection using a QBM, we restrict the data set to three dimensions. To satisfy (2) and (3) within the limitations of current hardware capabilities, we thus generate 1007 3D data points $x\in\left[0,...,127\right]^3$ clustered in five clusters and containing seven anomalies using the method \texttt{make\_blobs} from scikit-learn \cite{scikit-learn}. Therefore, seven bits are required per dimension, which means that 21 visible neurons are needed to represent the QBM's input, i.e., a single data point. Due to the $50/50$ train/test split, the ratio of anomalies is $\leq 7/500\approx 1\%$, satisfying (2). A pairplot of the generated data set is displayed in figure~\ref{fig:dataset}.

\subsection{Hyperparameter Optimization}
\label{subsec:hyperparameter}
Being generative models, the BMs have several hyperparameters which need to be optimized to achieve decent results. Choosing a greedy optimization strategy due to heavy computation time demands, we tuned the following hyperparameters descending in their typical relevance: (1) The number of hidden neurons, (2) the number of epochs and finally (3) the batchsize, while choosing standard values for the latter ones inspired by \cite{Amin18}. Notably, the learning rate was determined independent of all other hyperparameters in an empirical pre-study to the evaluation. For details on our implementation, see \url{https://github.com/jonas-stein/QBM-Anomaly-Detection}.


As we estimate that the here conducted hyperparameter search would take roughly two days of QPU time, our limited access to D-Wave's Annealers ($\sim$3 hours) necessitates a hyperparameter search using classical simulators for sampling. As proposed by \cite{QCE57702.2023.10182}, Simulated Annealing (SA) is a suitable choice for this, as it (like a quantum annealer \cite{PhysRevE.91.012104}) intrinsically approximates a Boltzmann distribution for a given temperature when using a suitable annealing schedule and neighbor generation (for details, see the articles of \cite{7782986,PhysRevX.6.031015}). 

\begin{figure}[ht]
    \centering
    \includegraphics[width=\columnwidth]{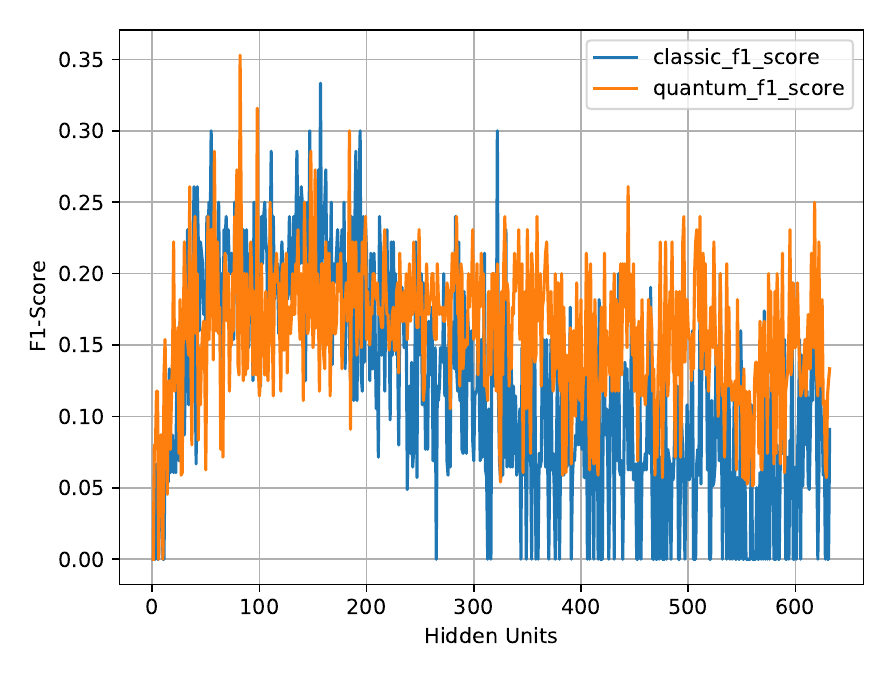}
    \caption{F1 score for increasing number of hidden neurons. The blue line shows the classical RBM and the orange line shows the QBM using the Simulated Annealing based sampler.}
    \label{fig:hidden_units}
\end{figure}

Starting our hyperparameter optimization with the number of hidden neurons, we respect the physically possible qubit embeddings on the employed D-Wave QPUs, which empirically restricts us to a maximum of 94 and respectively 632 hidden neurons for the \emph{D-Wave 2000Q} and the \emph{Advantage 4.1}. Figure~\ref{fig:hidden_units} illustrates that the QBM requires significantly fewer hidden neurons to reach its optimal F1 score compared to the RBM: While the QBMs optimum is at 82 hidden neurons with an F1 score of $0.35$, the RBM reaches its optimal F1 score of $0.33$ at 157 hidden neurons. Thus, the QBM achieves a better result with fewer resources. This is most likely the case, as the QBM can generally model complex dependencies better compared to a similarly sized RBM, as it allows for lateral connections. However, as the F1 scores show a large variance for small changes in the number of hidden units, caution has to be taken when concluding from these results, as it appears that the number of employed seeds (three) might be too low for undeniable statistical relevance.


\begin{figure}[ht]
    \centering
    \includegraphics[width=\columnwidth]{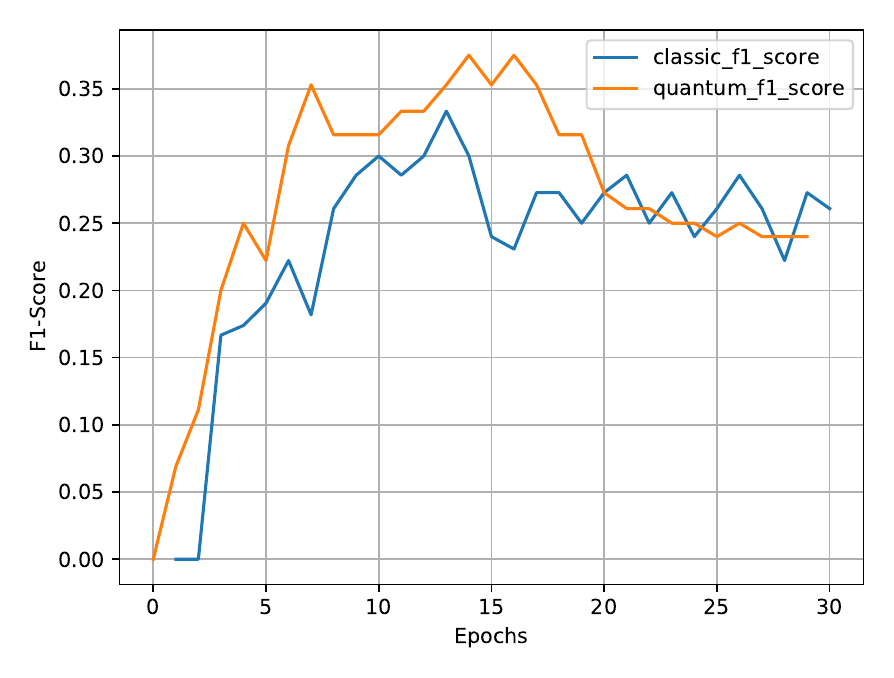}
    \caption{F1 score for increasing number of epochs. The blue line shows the classical RBM and the orange line shows the QBM using the Simulated Annealing based sampler.}
    \label{fig:epochs}
\end{figure}

Subsequently, the number of epochs was optimized, while each approach was fitted with the previously determined optimal number of hidden neurons. Figure~\ref{fig:epochs} displays the results which clearly show that the QBM again outperforms the RBM. While the RBM reaches its optimum at 13 epochs, with an F1 score of $0.33$, the optimum of the QBM is at 14 and 16 epochs, respectively, with an F1 score of $0.375$. In addition, the QBM consistently yields better results than the RBM approach at fewer epochs. At seven epochs, the F1 score of the QBM already reaches $0.35$, which exceeds the global optimum of the RBM which is obtained at 13 epochs. Thus, the QBM only needs about half the number of epochs to reach an even better performance that the RBM. Due to the limited available computing time on quantum hardware, seven epochs are are chosen for further optimization steps, and 13 for the RBM.
\begin{figure}[t]
    \centering
    \includegraphics[width=\columnwidth]{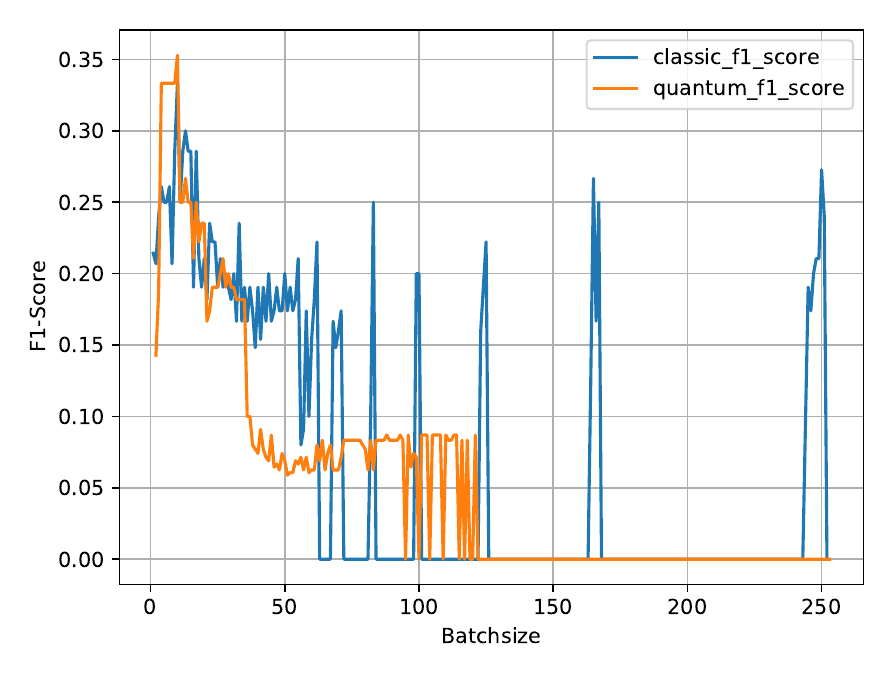}
    \caption{F1 score for increasing batch sizes. The blue line shows the classical RBM and the orange line shows the QBM using the Simulated Annealing based sampler.}
    \label{fig:batchsize}
\end{figure}

Concluding the hyperparameter optimization, Figure~\ref{fig:batchsize} shows that both the RBM and the QBM reach their optimum at a batch size of ten. The RBM remains at an F1 score of $0.33$, while the QBM achieves $0.35$.

\subsection{QPU Results}
Using the optimized hyperparameters, we now explore the result quality of the \emph{D-Wave 2000Q} and the \emph{Advantage 4.1} in comparison with the SA sampler and the RBM baseline. Table~\ref{fig:hardware_comparison} depicts the obtained results, which show that the Quantum Annealers are (still) outperformed by the classical approaches. Both quantum annealers missed one anomaly and thus achieved an identical recall. The \emph{Advantage 4.1} however identified less false positives according to its $62\%$ higher precision.
\begin{table}[h!]
    \centering
    \begin{tabular}{l | c | c | c | c }
     & RBM & SA & 2000Q & Advantage \\
     \hline
     F1-Score & 0.33 & 0.35 & 0.14 & 0.21 \\
     Recall & 1 & 1 & 0.67 & 0.67 \\
     Precision & 0.2 & 0.21 & 0.08 & 0.13 \\
    \end{tabular}
    \caption{Result quality of all examined approaches.}
    \label{fig:hardware_comparison}
\end{table}

\begin{figure}[t]
    \centering
    \includegraphics[width=\columnwidth]{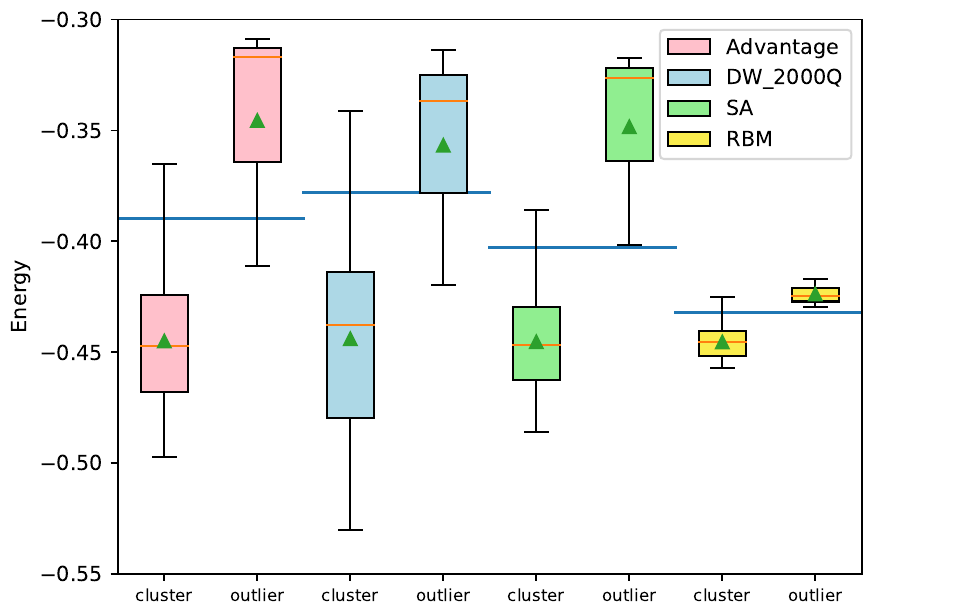}
    \caption{Normalized energies of anomalies and cluster points for different approaches. The blue line represents the chosen energy threshold separating normal from anomalous data.}
    \label{fig:Energies}
\end{figure}

Fig.~\ref{fig:Energies} shows a compact overview for comparing the energy levels in form of a box plot. Interestingly, the variance in the purely classical RBM approach is significantly lower than for all QBM variants, which might be caused by the less complex model, as it does not have any lateral neuron connections. Comparing the three QBM results, we clearly see that SA achieved the best performance, which makes sense, as the hyperparameters were trained for it. For the D-Wave QPUs a clear improvement can be observed for the newer \emph{Advantage 4.1} system compared to the older \emph{2000Q} model, which give the promising outlook of outperforming purely classical approaches with future hardware generations if this trend continues. We expect the main reasons for this to be the higher error rate of the \emph{2000Q} and the larger number of employed qubits, as its weaker connectivity demands for more ancillary qubits for the same number of hidden units.


\section{\uppercase{Conclusion}}
\label{sec:conclusion}
Our results indicate that QBMs can outperform their classically employed analog (RBMs) for unsupervised anomaly detection in terms of training steps and result quality. With limited access to quantum hardware however, we were unable to yield evidence for this claim when using Quantum Annealers as samplers, most probably because the hyperparameter tuning was conducted using a simulated annealing sampler that merely approximates the QPU results. Furthermore, the results show that the more recent D-Wave \emph{Advantage 4.1} QPU achieves significantly better performance than its predecessor \emph{D-Wave 2000Q}, even suggesting a possible quantum advantage in case that the hardware performance continues to improve similarly in the future. 

To improve the results of the quantum approach for future work, we suggest to implement a classical sampler that more closely matches the results of the utilized quantum hardware in a high performance computing oriented programming language to improve the accuracy and statistical relevance of the hyperparameter search. If successful, this should allow an upscaling to a more realistic dataset dimensionality to gradually approach the limitations of classical approaches for this task. This should also facilitate the usage of a dataset containing benign and malicious anomalies, to allow for comparing the results of the QBM with classical baselines in this regard. Furthermore, a closer evaluation in terms of training steps for a higher number of data points would be very interesting, as we expect our approach to be more efficient than the classical baselines in this regard, based on our experimental results.




\section*{\uppercase{Acknowledgements}}
This paper was partially funded by the German Federal Ministry for Economic Affairs and Climate Action through the funding program "Quantum Computing -- Applications for the industry" based on the allowance "Development of digital technologies" (contract number: 01MQ22008A).

\bibliographystyle{apalike}
{\small
\bibliography{main}}


\end{document}